\documentclass[epj]{svjour}
\usepackage{graphics}
\usepackage{psfig}
\begin{document}
\title{$SU(3)_c\otimes SU(4)_L\otimes U(1)_X$ model for three families}
%\subtitle{Do you have a subtitle?\\ If so, write it here}
\author{Luis A. S\'anchez\inst{1} \and Felipe A. P\'erez\inst{1}\and William A. Ponce\inst{2}}                     
\institute{Escuela de F\'\i sica, Universidad Nacional de Colombia,
A.A. 3840, Medell\'\i n, Colombia \and Instituto de F\'\i sica, Universidad de Antioquia,
A.A. 1226, Medell\'\i n, Colombia}
\date{Received: date / Revised version: date}
% The correct dates will be entered by Springer
%
\abstract{
An extension of the Standard Model to the local gauge group
$SU(3)_c\otimes SU(4)_L\otimes U(1)_X$ as a three-family model is
presented. The model does not contain exotic electric charges and we
obtain a consistent mass spectrum by introducing an anomaly-free discrete
$Z_2$ symmetry. The neutral currents coupled to all neutral vector bosons
in the model are studied. By using experimental results from the CERN LEP, SLAC Linear Collider and atomic parity violation we constrain the mixing angle between two of the neutral currents in the model and the mass of the additional neutral gauge bosons to be $-0.0032\leq\sin\theta\leq 0.0031$ and $0.67\ \hbox{TeV}\leq M_{Z_2} \leq 6.1$ TeV at $95 \%$ C.L., respectively. %
 } 
\maketitle
\section{Introduction}
\label{intro}
The Standard Model (SM), based on the local gauge
group $SU(3)_c\otimes SU(2)_L\otimes U(1)_Y$ \cite{sm}, can be extended in
several different ways: first, by adding new fermion fields (adding a
right-handed neutrino field constitute its simplest extension and has
profound consequences, as the implementation of the see-saw mechanism, and
the enlarging of the possible number of local abelian symmetries that can
be gauged simultaneously); second, by augmenting the scalar sector to more
than one Higgs representation, and third by enlarging the local gauge
group. In this last direction $SU(4)_L\otimes U(1)_X$ as a flavor group
has been considered before in the literature \cite{su4,pgus,little} which,
among its best features, provides with an alternative to the problem of
the number $N_f$ of families, in the sense that anomaly cancellation is
achieved when $N_f=N_c=3, \; N_c$ being the number of colors of $SU(3)_c$
(also known as QCD). Moreover, this gauge structure has been used recently
in order to implement the so-called little Higgs mechanism \cite{little}.

The analysis of the local gauge structure $SU(3)_c\otimes SU(4)_L\otimes U(1)_X$ (hereafter the 3-4-1 group) presented in the appendix of Ref.~\cite{pgus} shows that we may write the most general electric charge 
operator for this group as 
\begin{equation}\label{ch} Q=aT_{3L}+\frac{b}{\sqrt{3}}T_{8L}+
\frac{c}{\sqrt{6}}T_{15L}+ XI_4, \end{equation} 
where $a,b$ and $c$ are free parameters, 
$T_{iL}=\lambda_{iL}/2$, with $\lambda_{iL}$ the Gell-Mann matrices for
$SU(4)_L$ normalized as Tr$(\lambda_i\lambda_j)=2\delta_{ij}$, and
$I_4=Dg(1,1,1,1)$ is the diagonal $4\times 4$ unit matrix. The $X$ values 
are fixed by anomaly cancellation of the fermion content in 
the possible models and an eventual coefficient for $XI_4$ can be absorbed 
in the $X$ hypercharge definition. The free parameters $a,b$ and $c$ fix 
the gauge boson structure of the electroweak sector $[SU(4)_L\otimes 
U(1)_X]$, and also the electroweak charges of the scalar representations 
which are fully determined by the symmetry breaking pattern implemented. 
In particular $a=1$ gives the usual isospin of the known electroweak 
interactions, with $b$ and $c$ remaining as free parameters, producing an 
infinite plethora of possible models.

Restricting the particle content of the model to particles without exotic
electric charges and by paying due attention to anomaly cancellation, a
few different models are generated \cite{pgus}. In particular, the
restriction to ordinary electric charges, in the fermion, scalar and gauge
boson sectors, allows only for two different cases for the simultaneous
values of the parameters $b$ and $c$, namely: $b= c = 1$ and $b = 1, c =
-2$, which become a convenient classification scheme for these type of
models. Models in the first class differ from those in the second one not
only in their fermion content but also in their gauge and scalar boson
sectors. Four of the identified models without exotic electric charges are
three-family models in the sense that anomalies cancel among the three
families of quarks and leptons in a nontrivial fashion. Two of them are
models for which $b= c = 1$, and one of them has been analyzed in Ref.~\cite{pgus}. The other two models belong to the class for which $b = 1, c = -2$ and one of them, the so-called ``Model {\bf E}" in the appendix of Ref.~\cite{pgus}, will be studied in this paper. It is worth noticing that in the four different models at least one of the three families is treated differently.

This paper is organized as follows. In the next section we describe the
fermion content of the particular model we are going to study. In Sect.~\ref{sec:2} we introduce the scalar sector. In Sect.~\ref{sec:3}
we study the gauge boson sector paying special attention to the neutral currents present in the model and their mixing. In Sect.~\ref{sec:4}
we analyze the fermion mass spectrum. In Sect.~\ref{sec:5} we
use experimental results in order to constrain the mixing angle between
two of the neutral currents in the model and the mass scale of the new
neutral gauge bosons. In the last section we summarize the model and state
our conclusions.

\section{The Fermion Content of the Model}
\label{sec:1}
In what follows we assume that the electroweak gauge group is
$SU(4)_L\otimes U(1)_X$ which contains $SU(2)_L\otimes U(1)_Y$ as a
subgroup. We will consider the case of a non-universal hypercharge $X$ in
the quark sector, which implies anomaly cancellation among the three
families in a non-trivial fashion.

Here we are interested in studying the phenomenology of three-family
models without exotic electric charges and with values $b=1$, $c=-2$ for
the parameters in the electric charge generator in Eq.~(\ref{ch}). As an
example we take Model {\textbf E} of Ref.~\cite{pgus} for which the
electric charge operator is given by $Q=T_{3L}+T_{8L}/\sqrt{3}-
2T_{15L}/\sqrt{6}+ XI_4$. The model has the following anomaly free
fermion structure:

\[\begin{array}{ccccc}\hline
Q_{1L}=\left(\begin{array}{c}d_1\\u_1\\U_1\\D_1 
\end{array}\right)_L & 
d^c_{1L} & u^c_{1L}&
U^c_{1L}& D^{c}_{1L} \\ \hline [3,4^*,{1\over 6}] & 
[3^*,1,{1\over 3}] & [3^*,1,-{2\over 3}]
& [3^*,1,-{2\over 3}]& [3^*,1,{1\over 3}] \\ \hline
\end{array} \]

\[\begin{array}{ccccc}\hline
Q_{jL}=\left(\begin{array}{c}u_j\\d_j\\D_j\\U_j \end{array}\right)_L &
u^c_{jL} & d^c_{jL}& D^c_{jL}& U^{c}_{jL} \\ \hline [3,4,\frac{1}{6}] &
[3^*,1,-{2\over 3}] & [3^*,1,{1\over 3}] & [3^*,1,{1\over 3}]&
[3^*,1,-{2\over 3}] \\ \hline \end{array} \]

\[\begin{array}{ccc}\hline
L_{\alpha L}=\left(\begin{array}{c} e^-_\alpha\\ \nu_{e \alpha}\\ 
N^0_\alpha \\ E^{-}_\alpha \end{array}\right)_L & 
e^+_{\alpha L} & E^+_{\alpha L}\\ \hline
[1,4^*,-{1\over 2}] & [1,1,1] & [1,1,1]\\ \hline
\end{array} \]

\noindent where $j=2,3$ and $\alpha = 1,2,3$ are two and three family
indexes, respectively. The numbers in parenthesis refer to the $[SU(3)_C,
SU(4)_L, U(1)_X]$ quantum numbers, respectively. Notice that if needed,
the lepton structure of the model can be augmented with an undetermined
number of neutral Weyl singlet states $N^{0}_{L,n} \sim [1,1,0]$,
$n=1,2,...,$ without violating our assumptions, neither the anomaly
constraint relations, because singlets with no $X$-charges are as good as
not being present as far as anomaly cancellation is concerned.

\section{The Scalar Sector}
\label{sec:2}
Our aim is to break the symmetry following the pattern
\begin{eqnarray}\nonumber
SU(3)_c\otimes SU(4)_L\otimes & U(1)_X & \\* \nonumber  
& \rightarrow & SU(3)_c\otimes SU(3)_L\otimes
U(1)_X \\* \nonumber 
& \rightarrow & SU(3)_c\otimes SU(2)_L\otimes U(1)_Y \\* \nonumber 
& \rightarrow & SU(3)_c\otimes U(1)_Q, \nonumber
\end{eqnarray}
\noindent
where $SU(3)_c\otimes SU(3)_L\otimes U(1)_X$ refers to the so-called
3-3-1 structure introduced in Ref.~\cite{pfs}. At the same time we want to give masses to the fermion fields in the model.  With this in mind we introduce the following four Higgs scalars: 
$\phi_1[1,4^*,-1/2]$ with a Vacuum Expectation Value (VEV) aligned in the direction $\langle\phi_1\rangle=(0,v,0,0)^T$; $\phi_2[1,4^*,-1/2]$ 
with a VEV aligned as $\langle\phi_2\rangle=(0,0,V,0)^T$;
$\phi_3[1,4,-1/2]$ with a VEV aligned in the direction $\langle\phi_3\rangle=(v^\prime,0,0,0)^T$, and $\phi_4[1,4,-1/2]$ with a VEV aligned as $\langle\phi_4\rangle=(0,0,0,V^\prime)^T,$ with the
hierarchy $V\sim V^\prime >> \sqrt{v^2 + v^{\prime 2}} \simeq 174$~GeV (the electroweak breaking scale).

\section{The Gauge Boson Sector}
\label{sec:3}
In the model there are a total of 24 gauge bosons: One gauge field
$B^\mu$ associated with $U(1)_X$, the 8 gluon fields associated
with $SU(3)_c$ which remain massless after breaking the symmetry, and 
another 15 gauge fields associated with $SU(4)_L$ which, for $b=1$ and 
$c=-2$, can be written as 
\[{1\over 2}\lambda_\alpha A^\mu_\alpha={1\over \sqrt{2}}\left(
\begin{array}{cccc}D^\mu_1 & W^{+\mu} & K^{+\mu} & X^{0\mu}\\ W^{-\mu} &
D^\mu_2 &  K^{0\mu} &  X^{-\mu}\\
K^{-\mu} & \bar{K}^{0\mu} & D^\mu_3 & Y^{-\mu}\\
\bar{X}^{0\mu} & X^{+\mu} & Y^{+\mu} & D^\mu_4 \end{array}\right), \]
where $D^\mu_1=A_3^\mu/\sqrt{2}+A_8^\mu/\sqrt{6}+A_{15}^\mu/\sqrt{12},\;
D^\mu_2=-A_3^\mu/\sqrt{2}+A_8^\mu/\sqrt{6}+A_{15}^\mu/\sqrt{12}$;
$D^\mu_3=-2A_8^\mu/\sqrt{6}+A_{15}^\mu/\sqrt{12}$, and 
$D^\mu_4=-3 A_{15}^\mu/\sqrt{12}$.

After breaking the symmetry with $\langle\phi_1\rangle +
\langle\phi_2\rangle+ \langle\phi_3\rangle+ \langle\phi_4\rangle$ and 
using for the covariant derivative for 4-plets $iD^\mu=
i\partial^\mu-g\lambda_\alpha A^\mu_\alpha/2-g'XB^\mu$, 
where $g$ and $g^\prime$ are the $SU(4)_L$ and $U(1)_X$ gauge coupling 
constants respectively, we get the following mass terms for the charged 
gauge bosons: 
$M^2_{W^\pm}=g^2(v^2+v^{\prime 2})/2$, 
$M^2_{K^\pm}=g^2(v^{\prime 2}+V^2)/2$, 
$M^2_{X^\pm}=g^2(v^2+V^{\prime 2})/2$,
$M^2_{Y^\pm}=g^2(V^2+V^{\prime 2})/2$,
$M^2_{K^0(\bar{K}^0)}=g^2(v^2 + V^2)/2$, 
and $M^2_{X^0(\bar{X}^0)}=g^2(v^{\prime 2}+V^{\prime 2})/2$. 
Since $W^\pm$ does not mix with the other charged bosons we have that 
$\sqrt{v^2+v^{\prime 2}}\approx 174$ GeV as mentioned in the previous 
section.

For the four neutral gauge bosons we get mass terms of the form 

\begin{eqnarray*}
M&=&{g^2\over 2}\Big\{V^2 \left(\frac{g'B^\mu}{g}
-\frac{2A_8^\mu}{\sqrt{3}}+\frac{A^\mu_{15}}{\sqrt{6}}\right)^2 \\
& & + V^{\prime 2}\left(\frac{g'B^\mu}{g}+\frac{3 
A^\mu_{15}}{\sqrt{6}}\right)^2 \\
& &+v^{\prime 2}\left(A^\mu_3
+\frac{A_8^\mu}{\sqrt{3}}+\frac{A^\mu_{15}}{\sqrt{6}}
-\frac{g'B^\mu}{g}\right)^2 \\
& & + v^2 \left(\frac{g'B^\mu}{g}-A^\mu_3
+\frac{A_8^\mu}{\sqrt{3}}+\frac{A^\mu_{15}}{\sqrt{6}}\right)^2 \Big\}.
\end{eqnarray*}
\noindent 
$M$ is a $4 \times 4$ matrix with a zero eigenvalue
corresponding to the photon. Once the photon field has been identified, we
remain with a $3 \times 3$ mass matrix for three neutral gauge bosons
$Z^\mu$, $Z^{'\mu}$ and $Z^{''\mu}$. Since we are interested now in 
the low energy phenomenology of our model, we can choose $V=V^\prime$ in order to simplify matters. Also, the mixing between the three neutral gauge bosons can be further simplified by choosing $v^\prime = v$. For this particular case the field $Z''^\mu= 2A_8^\mu /\sqrt{6}+A_{15}^\mu/\sqrt{3}$ decouples from the other two and acquires a squared mass $(g^2/2)(V^2+v^2)$. By diagonalizing the remaining $2 \times 2$ mass matrix we get other two physical neutral gauge bosons which are defined through the mixing angle $\theta$ between $Z_\mu,\; Z'_\mu$ 

\begin{eqnarray}\nonumber
Z_1^\mu&=&Z_\mu \cos\theta+Z'_\mu \sin\theta \; ,\\ \nonumber
Z_2^\mu&=&-Z_\mu \sin\theta+Z'_\mu \cos\theta, \end{eqnarray} 
where
\begin{equation} \label{tan} 
\tan(2\theta) =  \frac{S_W^2 \sqrt{C_{2W}}}
{(1+S_W^2)^2 + \frac{V^2}{v^2}C_W^4 - 2}.
\end{equation}
\noindent 
$S_W= g^\prime /\sqrt{2 g^{\prime 2} + g^2}$ and $C_W$ are the sine 
and cosine of the electroweak mixing angle, respectively, and
$C_{2W}=C_W^2 - S_W^2$.

The photon field $A^\mu$ and the fields $Z_\mu$ and $Z'_\mu$ are given by

\begin{eqnarray} \nonumber
A^\mu&=&S_W A_3^\mu \nonumber \\
& & + C_W\left[\frac{T_W}{\sqrt{3}}\left(A_8^\mu-
2\frac{A_{15}^\mu}{\sqrt{2}}\right)+(1-T_W^2)^{1/2}B^\mu\right]\;,\nonumber \\  
Z^\mu&=& C_W A_3^\mu \nonumber \\
& & - S_W\left[\frac{T_W}{\sqrt{3}}\left(A_8^\mu-
2\frac{A_{15}^\mu}{\sqrt{2}}\right)+(1-T_W^2)^{1/2}B^\mu\right] \; , \nonumber \\ \label{fzzp}
Z'^\mu&=&\frac{1}{\sqrt{3}}(1-T_W^2)^{1/2}\left(A_8^\mu-
2\frac{A_{15}^\mu}{\sqrt{2}}\right)-T_W B^\mu.
\end{eqnarray}
\noindent
We can also identify the $Y$ hypercharge associated with the SM abelian gauge boson as
\begin{equation}\label{y}
Y^\mu=\frac{T_W}{\sqrt{3}}\left(A_8^\mu-
2\frac{A_{15}^\mu}{\sqrt{2}}\right)+(1-T_W^2)^{1/2}B^\mu.
\end{equation}

\subsection{Charged currents}
The Hamiltonian for the charged currents in the model is given by

\begin{eqnarray}\nonumber
H^{CC}&=&{g\over \sqrt{2}}\lbrace W^+_\mu [(\sum_{j=2}^3\bar{u}_{aL}\gamma^\mu 
d_{aL})- \bar{u}_{1L}\gamma^\mu d_{1L} \\ \nonumber & &
- (\sum_{\alpha=1}^3
\bar{\nu}_{e\alpha L}\gamma^\mu e^-_{\alpha L})] \\ \nonumber & &
+K^+_\mu[(\sum_{j=2}^3\bar{u}_{aL}\gamma^\mu 
D_{aL})-\bar{U}_{1L}\gamma^\mu d_{1L} \\ \nonumber & &
- (\sum_{\alpha=1}^3
\bar{N}^0_{\alpha L}\gamma^\mu e^-_{\alpha L})] \\ \nonumber & & 
+X^+_\mu[(\sum_{j=2}^3\bar{U}_{aL}\gamma^\mu 
d_{aL})-\bar{u}_{1L}\gamma^\mu D_{1L} \\ \nonumber & &
- (\sum_{\alpha=1}^3
\bar{\nu}_{e\alpha L}\gamma^\mu E^-_{\alpha L})] \\ \nonumber & &
+Y^+_\mu[(\sum_{j=2}^3\bar{U}_{aL}\gamma^\mu 
D_{aL})-\bar{U}_{1L}\gamma^\mu D_{1L} \\ \nonumber & &
- (\sum_{\alpha=1}^3
\bar{N}^0_{\alpha L}\gamma^\mu E^-_{\alpha L})] \\ \nonumber & & 
+K^0_\mu[(\sum_{j=2}^3\bar{d}_{aL}\gamma^\mu 
D_{aL})-\bar{U}_{1L}\gamma^\mu u_{1L} \\ \nonumber & &
- (\sum_{\alpha=1}^3
\bar{N}^0_{\alpha L}\gamma^\mu \nu_{e\alpha L})] \\ \nonumber & & 
+X^0_\mu[(\sum_{j=2}^3\bar{u}_{aL}\gamma^\mu 
U_{aL})-\bar{D}_{1L}\gamma^\mu d_{1L} \\ \nonumber & &
- (\sum_{\alpha=1}^3
\bar{E}^-_{\alpha L}\gamma^\mu e^-_{\alpha L})]\rbrace + H.c.
\end{eqnarray}
 
\subsection{Neutral currents}
The neutral currents $J_\mu(EM)$, $J_\mu(Z)$, $J_\mu(Z')$, and $J_\mu(Z'')$
associated with the Hamiltonian

\begin{eqnarray*}
H^0 &=& eA^\mu J_\mu(EM)+(g /{C_W})Z^\mu J_\mu(Z) \\
& &+ (g')Z'^\mu J_\mu(Z') + (g/(2\sqrt{2}))Z''^\mu J_\mu(Z''),
\end{eqnarray*}

are:

\begin{eqnarray}\nonumber
J_\mu(EM)&=&{2\over 3}\lbrack \sum_{j=2}^3(\bar{u}_a\gamma_\mu u_a+
\bar{U}_a\gamma_\mu U_a)
+\bar{u}_1\gamma_\mu u_1 \\ \nonumber
& & +\bar{U}_1\gamma_\mu U_1 \rbrack -{1\over3}\lbrack 
\sum_{j=2}^3(\bar{d}_a\gamma_\mu d_a+ \bar{D}_a\gamma_\mu D_a) \\ \nonumber 
& & +\bar{d}_1\gamma_\mu d_1+\bar{D}_1\gamma_\mu D_1 \rbrack \\ \nonumber
& & -\sum_{\alpha=1}^3\bar{e}^-_\alpha\gamma_\mu e^-_\alpha -\sum_{\alpha=1}^3\bar{E}^-_\alpha\gamma_\mu E^-_\alpha \\ \nonumber
&=&\sum_f q_f\bar{f}\gamma_\mu f,\\* \nonumber
J_\mu(Z)&=&J_{\mu,L}(Z)-S^2_WJ_\mu(EM),\\ \nonumber
J_\mu(Z')&=&J_{\mu,L}(Z')-T_WJ_\mu(EM),\\ \nonumber
J_\mu(Z'')&=&\sum_{a=2}^3(\bar{u}_{aL}\gamma_\mu u_{aL}+\bar{d}_{aL}\gamma_\mu d_{aL}-\bar{D}_{aL}\gamma_\mu 
D_{aL} \\ \nonumber
& & -\bar{U}_{aL}\gamma_\mu U_{aL})-\bar{d}_{1L}\gamma_\mu d_{1L}-\bar{u}_{1L}\gamma_\mu u_{1L} \\ \nonumber
& & +\bar{U}_{1L}\gamma_\mu U_{1L}+\bar{D}_{1L}\gamma_\mu D_{1L}\\ \nonumber
& & +\sum_{\alpha=1}^3(-\bar{e}^-_{\alpha L}\gamma_\mu e^-_{\alpha L}-\bar{\nu}_{e \alpha L}\gamma_\mu \nu_{e \alpha L}\\ 
& & +\bar{N}^0_{\alpha L}\gamma_\mu N^0_{\alpha L}
+\bar{E}^-_{\alpha L}\gamma_\mu E^-_{\alpha L}),
\end{eqnarray}
\noindent 
where $e=gS_W=g'C_W\sqrt{1-T_W ^2}>0$ is the electric charge, 
$q_f$ is the electric charge of the fermion $f$ in units of $e$, and 
$J_\mu(EM)$ is the electromagnetic current. Note from $J_\mu(Z'')$ that, notwithstanding the extra neutral gauge boson $Z''_\mu$ does not mix with $Z_\mu$ or $Z'_\mu$ (for the particular case $V=V^\prime$ and $v=v^\prime$), it still couples to ordinary fermions. 

The left-handed currents are
\begin{eqnarray} \nonumber
J_{\mu,L}(Z)&=&{1\over 2}[\sum_{j=2}^3(\bar{u}_{aL}\gamma_\mu u_{aL}
-\bar{d}_{aL}\gamma_\mu d_{aL})\\ \nonumber
& & -(\bar{d}_{1L}\gamma_\mu d_{1L}- \bar{u}_{1L}\gamma_\mu u_{1L})\\ 
\nonumber
& & -\sum_{\alpha=1}^3(\bar{e}^-_{\alpha L}\gamma_\mu e^-_{\alpha L}
-\bar{\nu}_{e\alpha L}\gamma_\mu \nu_{e\alpha L})] \\ \nonumber 
&=&\sum_f T_{4f}\bar{f}_L\gamma_\mu f_L ,
\end{eqnarray}

\begin{eqnarray}\nonumber
J_{\mu,L}(Z')&=&(2T_{W})^{-1}\lbrace\sum_{j=2}^3\lbrack 
T^2_W(\bar{u}_{aL}\gamma_\mu u_{aL}
-\bar{d}_{aL}\gamma_\mu d_{aL}) \\ \nonumber
& & -\bar{D}_{aL}\gamma_\mu D_{aL}+\bar{U}_{aL}\gamma_\mu U_{aL}\rbrack \\ \nonumber 
& & -T^2_W(\bar{d}_{1L}\gamma_\mu d_{1L}-\bar{u}_{1L}\gamma_\mu u_{1L}) \\ \nonumber
& & +\bar{U}_{1L}\gamma_\mu U_{1L}
-\bar{D}_{1L}\gamma_\mu D_{1L}\\ \nonumber 
& & +\sum_{\alpha=1}^3\lbrack -T^2_W(\bar{e}^-_{\alpha L}\gamma_\mu e^-_{\alpha L}- \bar{\nu}_{\alpha L}\gamma_\mu \nu_{\alpha L}) \\ \nonumber
& & +\bar{N}^0_{\alpha L}\gamma_\mu N^0_{\alpha L}-\bar{E}^-_{\alpha L}\gamma_\mu E^-_{\alpha L}\rbrack \rbrace
\\ &=&\sum_f
T'_{4f}\bar{f}_L\gamma_\mu f_L,
\end{eqnarray}
where $T_{4f}=Dg(1/2,-1/2,0,0)$ is the third component of the weak isospin
and $T'_{4f}=(1/2T_W)Dg(T^2_W, -T^2_W, -1, 1)$ = $T_W\lambda_3/2 +(1/T_W)(\lambda_8/(2\sqrt{3})-\lambda_{15}/\sqrt{6})$ is a convenient $4\times 4$ diagonal matrix, acting both of them on the representation 4 of $SU(4)_L$. Notice that $J_\mu(Z)$ is just the generalization of the neutral current present in the SM. This allows us to identify $Z_\mu$ as the neutral gauge boson of the SM, which is consistent with Eqs.~(\ref{fzzp}) and (\ref{y}).

The couplings of the mass eigenstates $Z_1^\mu$ and $Z_2^\mu$ are given by
\begin{eqnarray} \nonumber
H^{NC}&=&\frac{g}{2C_W}\sum_{i=1}^2Z_i^\mu\sum_f\{\bar{f}\gamma_\mu
[a_{iL}(f)(1-\gamma_5)\\ \nonumber & & 
+a_{iR}(f)(1+\gamma_5)]f\} \\ \nonumber
      &=&\frac{g}{2C_W}\sum_{i=1}^2Z_i^\mu\sum_f\{\bar{f}\gamma_\mu
      [g(f)_{iV}-g(f)_{iA}\gamma_5]f\},
\end{eqnarray}
where
\begin{eqnarray} \nonumber
a_{1L}(f)&=&\cos\theta(T_{4f}-q_fS^2_W)\\ \nonumber & &
+\frac{g'\sin\theta
C_W}{g} (T'_{4f}-q_fT_W)\;, \\ \nonumber
a_{1R}(f)&=&-q_fS_W\left(\cos\theta
S_W+\frac{g'\sin\theta}{g}\right)\;,\\ \nonumber
a_{2L}(f)&=&-\sin\theta(T_{4f}-q_fS^2_W)\\ \nonumber & &
+\frac{g'\cos\theta
C_W}{g} (T'_{4f}-q_fT_W)\;, \\ \label{a}
a_{2R}(f)&=&q_fS_W\left(\sin\theta S_W-\frac{g'\cos\theta}{g}\right),
\end{eqnarray}
and
\begin{eqnarray} \nonumber
g(f)_{1V}&=&\cos\theta(T_{4f}-2S_W^2q_f)\\ \nonumber & &
+\frac{g'\sin\theta}{g}
(T'_{4f}C_W-2q_fS_W)\;, \\ \nonumber
g(f)_{2V}&=&-\sin\theta(T_{4f}-2S_W^2q_f)\\ \nonumber & &
+\frac{g'\cos\theta}{g}
(T'_{4f}C_W-2q_fS_W) \;,\\ \nonumber g(f)_{1A}&=&\cos\theta
T_{4f}+\frac{g'\sin\theta}{g}T'_{4f}C_W\;, \\ \label{g}
g(f)_{2A}&=&-\sin\theta
T_{4f}+\frac{g'\cos\theta}{g}T'_{4f}C_W.
\end{eqnarray}
The values of $g_{iV},\; g_{iA}$ with $i=1,2$ are listed in Tables \ref{tab:1} and \ref{tab:2}.

\begin{table*}
\caption{The $Z_1^\mu\longrightarrow \bar{f}f$ couplings.}
\label{tab:1}
\begin{tabular}{lcc}
\hline\noalign{\smallskip}
$f$ & $g(f)_{1V}$ & $g(f)_{1A}$ \\ %\hline
\noalign{\smallskip}\hline\noalign{\smallskip}
$u_{1,2,3}$& $\cos\theta ({1\over 2}-{4S_W^2 \over 3})-
\frac{5\sin\theta}{6(C_{2W})^{1/2}}S_W^2$
& ${1\over 2} \cos\theta + \frac{\sin\theta}{2(C_{2W})^{1/2}}S_W^2$ \\ 
$d_{1,2,3}$ & $(-{1\over 2}+{2S_W^2\over 3})\cos\theta +\frac{\sin\theta}{6(C_{2W})^{1/2}}S^2_W$ 
& $-{1\over 2}\cos\theta - \frac{\sin\theta}{2(C_{2W})^{1/2}}S_W^2$ \\
$D_{1,2,3}$ & ${2S_W^2\over 3}\cos\theta +\frac{\sin\theta}{2(C_{2W})^{1/2}}({7 S_W^2 \over 3}-1)$ &
$-\frac{\sin\theta}{2(C_{2W})^{1/2}} C_W^2$ \\  
$U_{1,2,3}$ & $-{4S_W^2\over 3}\cos\theta-\frac{\sin\theta}{2(C_{2W})^{1/2}}({11 S_W^2 \over 3}-1)$ &
$\frac{\sin\theta}{2(C_{2W})^{1/2}} C_W^2$ \\
$e^-_{1,2,3}$& $\cos\theta (-{1\over 2}+2S_W^2)+ 
\frac{5\sin\theta}{2(C_{2W})^{1/2}}S_W^2 $ & 
$ -{\cos\theta\over 2} -\frac{\sin\theta}{2(C_{2W})^{1/2}}S_W^2$\\
$\nu_{1,2,3}$ &
${1\over 2}\cos\theta+ \frac{\sin\theta}{2(C_{2W})^{1/2}}S_W^2$ &
$ {1\over 2}\cos\theta+ \frac{\sin\theta}{2(C_{2W})^{1/2}}S_W^2$ \\ 
$N^0_{1,2,3}$ & $\frac{\sin\theta}{2(C_{2W})^{1/2}}C_W^2$ &
$\frac{\sin\theta}{2(C_{2W})^{1/2}}C_W^2$ \\ 
$E^-_{1,2,3}$ &
$2 S^2_W \cos\theta +\frac{\sin\theta}{(C_{2W})^{1/2}}(2-{5\over 2}C_W^2)$ & $-\frac{\sin\theta}{2(C_{2W})^{1/2}}C_W^2$ \\ 
\noalign{\smallskip}\hline
\end{tabular}
\end{table*}

\begin{table*}
\caption{The $Z_2^\mu\longrightarrow \bar{f}f$ couplings.}
\label{tab:2}
\begin{tabular}{lcc}
\hline\noalign{\smallskip}
$f$ & $g(f)_{2V}$ & $g(f)_{2A}$ \\ %\hline
\noalign{\smallskip}\hline\noalign{\smallskip}
$u_{1,2,3}$& $-\sin\theta ({1\over 2}-{4S_W^2 \over 3})-
\frac{5\cos\theta}{6(C_{2W})^{1/2}}S_W^2$
& $-{1\over 2} \sin\theta + \frac{\cos\theta}{2(C_{2W})^{1/2}}S_W^2$ \\ 
$d_{1,2,3}$ & $({1\over 2}-{2S_W^2\over 3})\sin\theta +\frac{\cos\theta}{6(C_{2W})^{1/2}}S^2_W$ 
& ${1\over 2}\sin\theta - \frac{\cos\theta}{2(C_{2W})^{1/2}}S_W^2$ \\
$D_{1,2,3}$ & $-{2S_W^2\over 3}\sin\theta +\frac{\cos\theta}{2(C_{2W})^{1/2}}({7 S_W^2 \over 3}-1)$ &
$-\frac{\cos\theta}{2(C_{2W})^{1/2}} C_W^2$ \\  
$U_{1,2,3}$ & ${4S_W^2\over 3}\sin\theta-\frac{\cos\theta}{2(C_{2W})^{1/2}}({11 S_W^2 \over 3}-1)$ &
$\frac{\cos\theta}{2(C_{2W})^{1/2}} C_W^2$ \\
$e^-_{1,2,3}$& $\sin\theta ({1\over 2}-2S_W^2)+ 
\frac{5\cos\theta}{2(C_{2W})^{1/2}}S_W^2 $ & 
$ {\sin\theta\over 2} -\frac{\cos\theta}{2(C_{2W})^{1/2}}S_W^2$\\
$\nu_{1,2,3}$ &
$-{1\over 2}\sin\theta+ \frac{\cos\theta}{2(C_{2W})^{1/2}}S_W^2$ &
$-{1\over 2}\sin\theta+ \frac{\cos\theta}{2(C_{2W})^{1/2}}S_W^2$ \\ 
$N^0_{1,2,3}$ & $\frac{\cos\theta}{2(C_{2W})^{1/2}}C_W^2$ &
$\frac{\cos\theta}{2(C_{2W})^{1/2}}C_W^2$ \\ 
$E^-_{1,2,3}$ &
$-2 S^2_W \sin\theta +\frac{\cos\theta}{(C_{2W})^{1/2}}(2-{5\over 2}C_W^2)$ & $-\frac{\cos\theta}{2(C_{2W})^{1/2}}C_W^2$ \\
\noalign{\smallskip}\hline
\end{tabular}
\end{table*}

As we can see, in the limit $\theta=0$ the couplings of $Z_1^\mu$ to 
the ordinary leptons and quarks are the same as in the SM; due to 
this we can test the new physics beyond the SM predicted by this 
particular model.

\section{Fermion Masses}
\label{sec:4}
The Higgs scalars introduced in Sect.~\ref{sec:2} break the 
symmetry in an appropriate way. Now, in order to generate both a simple mass splitting between ordinary and exotic fermions and a consistent mass spectrum, we introduce an anomaly-free discrete $Z_2$ symmetry \cite{KW}, with the following assignments of $Z_2$ charge $q$ 
\begin{eqnarray} \nonumber
q(Q_{aL}, u^c_{aL}, d^c_{aL}, L_{aL}, e^c_{aL}, \phi_1, \phi_3)&=& 0, \\ \label{z2d}
q(U^c_{aL}, D^c_{aL}, E^c_{aL}, \phi_2, \phi_4)&=& 1.
\end{eqnarray}
Notice that ordinary fermions are not affected by this discrete symmetry.

The gauge invariance and the $Z_2$ symmetry allow for the following Yukawa lagrangians:

\begin{itemize}
\item For quarks:
\begin{eqnarray} \nonumber
{\cal L}^Q_Y&=& \sum_{j=2}^3Q^T_{aL}C \Big\{\phi^*_3 \sum_{\alpha=1}^3 
h^u_{j\alpha}u^c_{\alpha L} + \phi^*_4 \sum_{\alpha=1}^3 
h^U_{j\alpha}U^c_{\alpha L} \\ \nonumber 
&+& \phi_1\sum_{\alpha=1}^3h^d_{j\alpha} d^c_{\alpha L} + \phi_2 
\sum_{\alpha=1}^3h^D_{j\alpha}D^c_{\alpha L} \Big\}
\\ \nonumber 
&+& Q^T_{1L}C \Big\{\phi^*_1\sum_{\alpha=1}^3 h^u_{1\alpha}u^c_{\alpha L} + \phi^*_2
\sum_{\alpha=1}^3h^U_{1 \alpha}U^c_{\alpha L} 
\\ \nonumber
&+& \phi_3 \sum_{\alpha=1}^3 
h^d_{1\alpha}d^c_{\alpha L} + \phi_4 \sum_{\alpha=1}^3 
h^D_{1\alpha}D^c_{\alpha L} \Big\}+ h.c., 
\end{eqnarray}
where the $h's$ are Yukawa couplings and $C$ is the charge conjugate operator.\\ 

\item For charged leptons:

\[{\cal L}_Y^l=\sum_{\alpha=1}^3\sum_{\beta=1}^3L_{\alpha L}^TC \Big\{\phi_3 h_{\alpha\beta}^e e_{\beta L}^+ +\phi_4h_{\alpha\beta}^E E_{\beta L}^+ \Big\}+ h.c.,\] 
\end{itemize}
\noindent
The Lagrangian ${\cal L}^Q_Y$ produces for up- and down-type quarks, in the basis $(u_1,u_2,u_3,U_1,U_2,U_3)$ and $(d_1,d_2,d_3,D_1,D_2,D_3)$ respectively, $6 \times 6$ block diagonal mass matrices of the form:

\[M_{uU}=\left(\begin{array}{cc}
M_{u(3\times 3)} &  0\\ 
0 &  M_{U(3\times 3)}\end{array}\right), \] 
where

\[M_u=\left(\begin{array}{ccc}
h^u_{11}v & h^u_{21}v^\prime & h^u_{31}v^\prime \\ 
h^u_{12}v & h^u_{22}v^\prime & h^u_{32}v^\prime \\ 
h^u_{13}v & h^u_{23}v^\prime & h^u_{33}v^\prime \end{array}\right),\] 

\[M_U=\left(\begin{array}{ccc}
h^U_{11}V & h^U_{21}V^\prime & h^U_{31}V^\prime\\  
h^U_{12}V & h^U_{22}V^\prime & h^U_{32}V^\prime\\  
h^U_{13}V & h^U_{23}V^\prime & h^U_{33}V^\prime\end{array}\right),\]
and

\[M_{dD}=\left(\begin{array}{cc}
M_{d(3\times 3)} &  0\\
0 &  M_{D(3\times 3)}\end{array}\right),\]
where

\[M_d=\left(\begin{array}{ccc}
h^d_{11}v^\prime & h^d_{21}v & h^d_{31}v \\
h^d_{12}v^\prime & h^d_{22}v & h^d_{32}v \\
h^d_{13}v^\prime & h^d_{23}v & h^d_{33}v \end{array}\right),\]

\[M_D=\left(\begin{array}{ccc}
h^D_{11}V^\prime & h^D_{21}V & h^D_{31}V\\
h^D_{12}V^\prime & h^D_{22}V & h^D_{32}V\\
h^D_{13}V^\prime & h^D_{23}V & h^D_{33}V\end{array}\right).\]

For the charged leptons the Lagrangian ${\cal L}_Y^l$, in the basis $(e_1,e_2,e_3,E_1,E_2,E_3)$, also produces a block diagonal mass matrix 

\[M_{eE}=\left(\begin{array}{cc}
M_{e(3\times 3)} &  0\\
0 &  M_{E(3\times 3)}\end{array}\right),\]
where the entries in the submatrices are given by
\begin{eqnarray}\nonumber
M_{e, \alpha \beta} = h^e_{\alpha \beta} v^\prime \qquad \mbox{and} & \qquad
M_{E, \alpha \beta} = h^E_{\alpha \beta} V^\prime.
\end{eqnarray}

The former mass matrices exhibit the mass splitting between ordinary and
exotic charged fermions and show that all the charged fermions in the
model acquire masses at the three level. Clearly, by a judicious tuning of
the Yukawa couplings and of the mass scales $v$ and $v^\prime$, a
consistent mass spectrum in the ordinary charged sector can be obtained.
In the exotic charged sector all the particles acquire masses at the scale
$V \sim V^\prime \gg 174$~GeV. Note that in the low energy limit our
model corresponds to a Type III two Higgs doublet model \cite{wsh} in which both doublets couple to the same type of fermions, with the quark and
lepton couplings treated asymmetrically.

The neutral leptons remain massless as far as we use only the original
fields introduced in Sect.~\ref{sec:1}. But as mentioned earlier, we may introduce
one or more Weyl singlet states $N^0_{L,b},\; b=1,2,....,$ which may
implement the appropriate neutrino oscillations \cite{neutrinos}.

\section{Constrains on the $(Z^\mu-Z'^{\mu})$ Mixing Angle and the 
$Z^{\mu}_2$ Mass}
\label{sec:5} 
To bound $\sin\theta$ and $M_{Z_2}$ we use parameters
measured at the $Z$ pole from CERN $e^+e^-$ collider (LEP), SLAC Linear 
Collider (SLC), and atomic parity violation constraints which are given
in Table \ref{tab:3}. 

The expression for the partial decay width for $Z^{\mu}_1\rightarrow
f\bar{f}$ is
 
\begin{eqnarray}\nonumber
\Gamma(Z^{\mu}_1\rightarrow f\bar{f})&=&\frac{N_C G_F
M_{Z_1}^3}{6\pi\sqrt{2}}\rho \Big\{\frac{3\beta-\beta^3}{2}
[g(f)_{1V}]^2 \\ \label{ancho}
& + & \beta^3[g(f)_{1A}]^2 \Big\}(1+\delta_f)R_{EW}R_{QCD}, \quad
\end{eqnarray}
\noindent 
where $f$ is an ordinary SM fermion, $Z^\mu_1$ is the physical gauge boson
observed at LEP, $N_C=1$ for leptons while for quarks
$N_C=3(1+\alpha_s/\pi + 1.405\alpha_s^2/\pi^2 - 12.77\alpha_s^3/\pi^3)$,
where the 3 is due to color and the factor in parenthesis represents the
universal part of the QCD corrections for massless quarks 
(for fermion mass effects and further QCD corrections which are 
different for vector and axial-vector partial widths see 
Ref.~\cite{kuhn}); $R_{EW}$ are the electroweak corrections which include the leading order QED corrections given by $R_{QED}=1+3\alpha/(4\pi)$. $R_{QCD}$ are further QCD corrections (for a comprehensive review see Ref.~\cite{leike} and references therein), and $\beta=\sqrt{1-4 m_f^2/M_{Z_1}^2}$ is a kinematic factor which can be taken equal to $1$ for all the SM fermions except for the bottom quark. 
The factor $\delta_f$ contains the one loop vertex
contribution which is negligible for all fermion fields except for the 
bottom quark for which the contribution coming from the top quark at the 
one loop vertex radiative correction is parametrized as $\delta_b\approx 
10^{-2} [-m_t^2/(2 M_{Z_1}^2)+1/5]$ \cite{pich}. The $\rho$ parameter 
can be expanded as $\rho = 1+\delta\rho_0 + \delta\rho_V$ where the 
oblique correction $\delta\rho_0$ is given by
$\delta\rho_0\approx 3G_F m_t^2/(8\pi^2\sqrt{2})$, and $\delta\rho_V$ is 
the tree level contribution due to the $(Z_{\mu} - Z'_{\mu})$ mixing which 
can be parametrized as $\delta\rho_V\approx
(M_{Z_2}^2/M_{Z_1}^2-1)\sin^2\theta$. Finally, $g(f)_{1V}$ and $g(f)_{1A}$
are the coupling constants of the physical $Z_1^\mu$ field with ordinary
fermions which are listed in Table \ref{tab:1}.

In what follows we are going to use the experimental values \cite{pdg}:
$M_{Z_1}=91.188$ GeV, $m_t=174.3$ GeV, $\alpha_s(m_Z)=0.1192$, 
$\alpha(m_Z)^{-1}=127.938$, and
$\sin^2\theta_W=0.2333$. The experimental values are introduced using the
definitions $R_\eta\equiv \Gamma(\eta\eta)/\Gamma(hadrons) $ for
$\eta=e,\mu,\tau,b,c$.

As a first result notice from Table \ref{tab:1}, that our model predicts 
$R_e=R_\mu=R_\tau$, in agreement with the experimental results in Table \ref{tab:3}.

The effective weak charge in atomic parity violation, $Q_W$, can be 
expressed as a function of the number of protons $(Z)$ and the number of 
neutrons $(N)$ in the atomic nucleus in the form 

\begin{equation}
Q_W=-2\left[(2Z+N)c_{1u}+(Z+2N)c_{1d}\right], 
\end{equation}
\noindent
where $c_{1q}=2g(e)_{1A}g(q)_{1V}$. The theoretical value for $Q_W$ for 
the Cesium atom is given by \cite{sirlin} $Q_W(^{133}_{55}Cs)=-73.09\pm0.04 
+ \Delta Q_W$, where the contribution of new physics is included in $\Delta 
Q_W$ which can be written as \cite{durkin}

\begin{equation}\label{DQ} 
\Delta 
Q_W=\left[\left(1+4\frac{S^4_W}{1-2S^2_W}\right)Z-N\right]\delta\rho_V
+\Delta Q^\prime_W.
\end{equation}

The term $\Delta Q^\prime_W$ is model dependent and it can be obtained 
for our model by using $g(e)_{iA}$ and $g(q)_{iV}$, $i=1,2$, from Tables \ref{tab:1} and \ref{tab:2}. The value we obtain is

\begin{equation}
\Delta Q_W^\prime=(3.75 Z + 2.56 N) \sin\theta + (1.22 Z + 0.41 N)
\frac{M^2_{Z_1}}{M^2_{Z_2}}\; .
\end{equation}

The discrepancy between the SM and the experimental data for $\Delta Q_W$ 
is given by \cite{casal}

\begin{equation}
\Delta Q_W=Q^{exp}_W-Q^{SM}_W=1.03\pm 0.44,
\end{equation}
which is $2.3\; \sigma$ away from the SM predictions.

\begin{table}
\caption{Experimental data and SM values for the parameters.}
\label{tab:3}
\begin{tabular}{lcl}
\hline\noalign{\smallskip}
& Experimental results & SM \\ %\hline
\noalign{\smallskip}\hline\noalign{\smallskip}
$\Gamma_Z$(GeV)  & $2.4952 \pm 0.0023$  &  $2.4966 \pm 0.0016$  \\   
$\Gamma(had)$ (GeV)  & $1.7444 \pm 0.0020$ & $1.7429 \pm 0.0015$ \\ 
$\Gamma(l^+l^-)$ (MeV) & $83.984\pm 0.086$ & $84.019 \pm 0.027$ \\
$R_e$ & $20.804\pm 0.050$ & $20.744\pm 0.018$ \\ 
$R_\mu$ & $20.785\pm 0.033$ & $20.744\pm 0.018$ \\ 
$R_\tau$ & $20.764\pm 0.045$ & $20.790\pm 0.018$ \\ 
$R_b$ & $0.21664\pm 0.00068$ & $0.21569\pm 0.00016$ \\ 
$R_c$ & $0.1729\pm 0.0032$ & $0.17230\pm 0.00007$ \\ 
$Q_W^{Cs}$ & $-72.65\pm 0.28\pm 0.34$ & $-73.10\pm 0.03$ \\
$M_{Z_{1}}$(GeV) & $ 91.1872 \pm 0.0021 $ & $ 91.1870 \pm 0.0021 $ \\ 
\noalign{\smallskip}\hline
\end{tabular}
\end{table}

\begin{figure*}
%\vspace*{11cm}       % Give the correct figure height in cm
%\resizebox{0.89\textwidth}{!}{
  \includegraphics{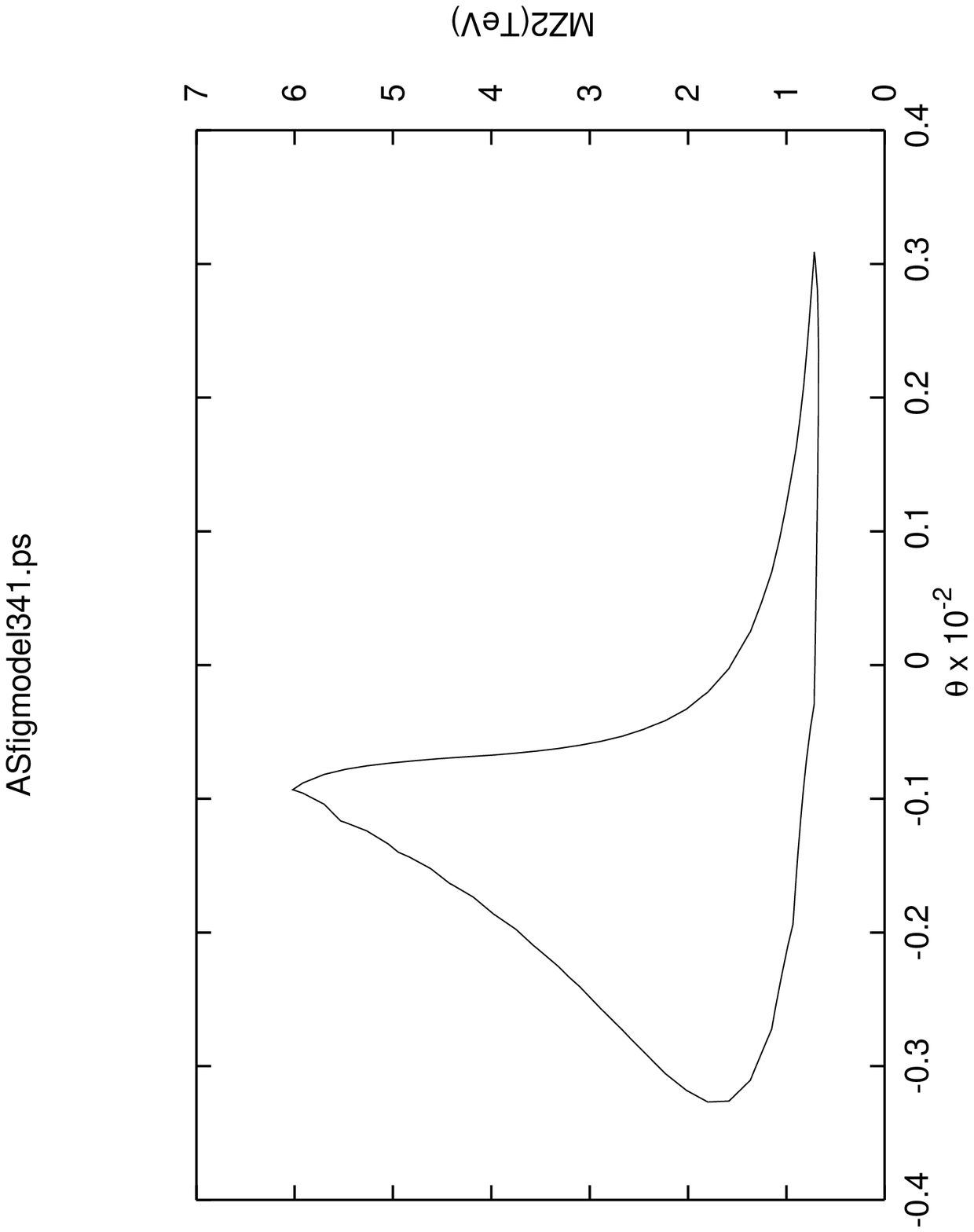}
%}
\caption{Contour plot displaying the allowed region for
$\theta$ vs. $M_{Z_2}$ at $95\%$ C.L.}
\label{fig:1}       % Give a unique label
\end{figure*}

Introducing the expressions for $Z$ pole observables in Eq.(\ref{ancho}),
with $\Delta Q_W$ in terms of new physics in Eq.(\ref{DQ}) and using
experimental data from LEP, SLC and atomic parity violation (see Table \ref{tab:3}), we do a $\chi^2$ fit and we find the best allowed region in the $(\theta-M_{Z_2})$ plane at $95\%$ confidence level (C.L.). In Fig.
\ref{fig:1} we display this region which gives us the constraints
\begin{equation} 
-0.0032\leq\theta\leq 0.0031, \;\;\; 0.67\; {\mbox TeV} 
\leq M_{Z_2} \leq 6.1\; {\mbox TeV}.
\end{equation} 

As we can see the mass of the new neutral gauge boson is compatible with
the bound obtained in $p\bar{p}$ collisions at the Fermilab Tevatron
\cite{abe}. 

%\begin{figure*}
%%\vspace*{11cm}       % Give the correct figure height in cm
%\resizebox{0.89\textwidth}{!}{
%  \includegraphics{fig1.ps}
%}
%\caption{Contour plot displaying the allowed region for
%$\theta$ vs. $M_{Z_2}$ at $95\%$ C.L.}
%\label{fig:1}       % Give a unique label
%\end{figure*}
 
\section{Conclusions} 
We have presented an anomaly-free model based on the local gauge group
$SU(3)_c\otimes SU(4)_L\otimes U(1)_X$, which does not contain exotic
electric charges. This last constraint fixes the values $b=1$ and $c=-2$ for the parameters in the electric charge generator in Eq.~(\ref{ch}).

We break the gauge symmetry down to $SU(3)_c\otimes U(1)_{Q}$ in an
appropriate way by using four different Higgs scalars $\phi_i,\:
i=1,2,3,4,$ which set two different mass scales: $V\sim V^\prime >>
\sqrt{v^2+v^{\prime 2}}\simeq 174$ GeV, with $v\sim v^\prime$. By
introducing an anomaly-free discrete $Z_2$ symmetry we also obtain a
simple mass splitting between exotic and ordinary fermions, and a
consistent mass spectrum both in the quark and in the lepton sector.
Notice also the consistence of our model in the charged lepton sector
where it predicts the correct ratios $R_\eta$, $\eta=e,\mu,\tau$, in the
$Z$ decays. This is a characteristic feature of the two class of three-family models introduced in Ref.~\cite{pgus}.

By using experimental results we obtain a lowest bound of $0.67$ TeV $\leq
M_{Z_2}$ for the mass of an extra neutral gauge boson $Z_2$, and we bound
the mixing angle $\theta$ between the SM neutral current and the $Z_2$ one
to be $-0.0032<\theta<0.0031$.

When we compare the numerical results presented in the previuos section with the results presented in Ref.~\cite{pgus}, we find that the mixing angle $\theta$ is of the same order of magnitude ($\sim 10^{-3}$), but for the model considered here the mass associated with the new neutral current has smaller lower and upper bounds, with the lower bound just below the TeV scale, which allows for a possible signal at the Fermilab Tevatron.

For our analysis we have choosen just one of the two possible three-family models without exotic electric charges, characterized by the parameters $b=-c/2=1$ in the electric charge operator \cite{pgus}. We believe that the low energy phenomenology for the other model must produce similar results than the ones presented in this paper.

\section*{Acknowledgments}
We akcnowledge partial financial support from Universidad Nacional de
Colombia-Sede Medell\'\i n, and from Universidad de Antioquia. We also 
thank Diego A. Guti\'errez for helping us with the numerical analysis 
presented in Sect.~\ref{sec:5}.

\end{document}